\documentclass[10pt,a4paper]{article}
\usepackage{url}
\usepackage{graphicx} 
\newcommand{\mean}[1]{\left\langle #1 \right\rangle} 
\newcommand{\abs}[1]{\left| #1 \right|} 

\renewcommand{\epsilon}{\varepsilon} 
\usepackage{fancyhdr}
\begin{document}

\makeatletter
\def\@biblabel#1{[#1]}
\makeatother

\markboth{Mario V. Tomasello, Claudio J. Tessone and Frank Schweitzer}{A model of dynamic rewiring and knowledge exchange\\ in R\&D networks}

%\catchline{}{}{}{}{}
%
%%%%%%%%%%%%%%%%%%%%%%%%%%%%%%%%%%%%%%%%%%%%%%%%%%%%%%%%%%%%%%%%%%%%

\title{\textbf{A model of dynamic rewiring and knowledge exchange in R\&D networks}}

\author{Mario V. Tomasello$^{*,1}$, Claudio J. Tessone$^{1,2}$ and Frank Schweitzer$^{1}$\\
{\footnotesize $^\ast$Corresponding author; E-mail: mtomasello@ethz.ch}\\
{\footnotesize $^1$Chair of Systems Design, ETH Zurich, Department of Management,}\\
{\footnotesize Technology and Economics, Weinbergstrasse 56/58, CH-8092 Zurich, Switzerland}\\
{\footnotesize $^2$URPP Social Networks, Department of Business Administration,}\\
{\footnotesize Universit\"at Z\"urich, Andreasstrasse 15, CH-8050 Zurich, Switzerland}
}

\date{ \footnotesize{Received: 7 May 2015 \\
Revised: 18 May 2016\\
Accepted: 19 May 2016\\
Published: 17 June 2016} \\
\vspace {10pt}
\normalsize{DOI: 10.1142/S0219525916500041}\\
}

\maketitle

\begin{abstract}

This paper investigates the process of knowledge exchange in inter-firm Research and Development (R\&D) alliances by means of an agent-based model.
Extant research has pointed out that firms select alliance partners considering both network-related and network-unrelated features (e.g., social capital versus complementary knowledge stocks).
In our agent-based model, firms are located in a metric knowledge space. The interaction rules incorporate an exploration phase and a knowledge transfer phase, during which firms search for a new partner and then evaluate whether they can establish an alliance to exchange their knowledge stocks.
The model parameters determining the overall system properties are the rate at which alliances form and dissolve and the agents' interaction radius.
Next, we define a novel indicator of performance, based on the distance traveled by the firms in the knowledge space. Remarkably, we find that -- depending on the alliance formation rate and the interaction radius -- firms tend to cluster around one or more attractors in the knowledge space, whose position is an emergent property of the system.
And, more importantly, we find that there exists an inverted U-shaped dependence of the network performance on both model parameters.

\end{abstract}

\vspace{12pt}

\normalsize

\section{Introduction}
\label{sec:intro}

%The realm of research of Physics has greatly expanded in the last years, encompassing also areas that were usually deprived of quantitative methods, or that relied on mainly theoretical constructions without proper accounting of the empirical observations. 
%For instance, some tools developed in Statistical Physics (for disordered and out-of-equilibrium systems) can be applied to social and economic systems, composed by many interacting and heterogenous agents.
%However, a non-negligible part of the Physics research in those areas is based mostly on speculative grounds: it is possible to find modeling approaches to social or economic interactions that are purely based on plausibility arguments.
%Our work contributes indeed to this strand of literature, by studying the exchange of knowledge in strategic Research and Development (R\&D) alliances, as well as the emergence of network structures and their features.

The number of observed inter-organizational Research and Development (R\&D) alliances has grown until the 1990s, especially in industrial sectors like IT, Pharmaceuticals and other high-technology ones \cite{ahuja2000collaboration,hagedoorn2002inter}.
This has stimulated research in different domains, for instance the mechanisms behind the formation of R\&D alliances \cite{powell2005network}, the complex networks they origin \cite{rosenkopf2007comparing,tomasello2014therole}, and the way they can be described, modeled and their evolution forecasted \cite{konig2012efficiency,garas2014selection}. 
A number of theoretical works have shown that, among many other reasons, there exist three main motives for firms to engage in alliances. 
First, they can gain access to different assets more quickly than they could do \textit{in-house} \cite{liebeskind96:_knowl_strat_theor_firm,das2000resource}.
Second, alliances foster the exchange of knowledge between firms: by joining their technological resources, firms can actually enlarge their knowledge basis more than they could do individually \cite{baum2000don,mowery1998technological,rosenkopf2003overcoming}. 
Third, firms can share the costs and risks of a project, especially when this is expensive or with uncertain outcome \cite{hagedoorn00:_resear}. 
All of these aspects -- even when a knowledge transfer is not directly involved -- result in a learning process by the firms \cite{ahuja2000duality}, of tacit or explicit knowledge, thus making R\&D alliances an important element in many firms' strategy.

In this work we investigate such a learning process, which we model as a knowledge exchange occurring after the establishment of an alliance between two agents. Our agents can change partners and rewire their links as well, thus introducing complex mutual feedbacks between the network structure and their intrinsic characteristics (i.e. their knowledge basis).

Our agent-based model follows an existing stream of literature in the direction of bounded confidence and continuous opinion dynamics models \cite{axelrod1997dissemination,deffuant2000mixing,degroot1974reaching,hegselmann2002opinion,groeber2009groups}, especially applied to innovation networks \cite{fischer2001knowledge,baum2010network}. In the wake of this previous work, we assume that the collaborating nodes are endowed with an evolving knowledge basis, that affects alliances and -- in its turn -- is affected by them. However, differently from the studies that have been done so far, our model does not focus on the formation of consensus clusters (see \cite{axelrod1997dissemination,schweitzer2009nonlinear} in the case of social systems, or \cite{fagiolo2003exploitation} for technology islands).
Also, our work differs from previous studies \cite{gersbach2003endogenous,suzumura1992cooperative} that are focused on strategic decisions made by firms and the effects that these have on the innovation incentives for the involved parties.
We rather focus on the dynamics that leads the system to the observed final state, with emphasis on the exploration of the knowledge space by the collaborating agents. We then investigate the existence of an optimal network dynamics that maximizes such a knowledge space exploration.

With respect to R\&D networks, it has been shown \cite{rosenkopf2007comparing} that -- despite long-term simultaneous fluctuations \cite{tomasello2013riseandfall} -- different industrial sectors exhibit different characteristics in their alliance activity (size and density of the corresponding inter-firm network, heterogeneity of degree distributions, other sophisticated topological network properties and so on).
Part of these observed differences have been explained with the so-called ``technological regime'' of the sector \cite{rosenkopf2007comparing}.
A technological regime is defined \cite{nelson82} as the pattern of behaviors and common practices in an industrial sector, that are influenced by factors such as technological dynamism, technological uncertainty or separability of innovation activities.
In the literature, two technological regimes have originally been detected \cite{winter84:_schum}: an \textit{entrepreneurial regime}, where R\&D activities are mainly carried out by new innovative firms, and a \textit{routinized regime}, where innovation is mainly done by incumbent firms.
These two extremes are often referred to as \textit{tacit knowledge regime} and \textit{explicit knowledge regime}, respectively, because firms in the network tend to interact with similar or with diverse firms (in terms of knowledge basis), in the respective cases.
However, this distinction has been extended over the years, leading to the identification of several classes of technological regimes, spanning between the two aforementioned extremes.

To the best of our knowledge, there is only little research about the influence of technological regimes on the formation of alliances, from a modeling point of view.
The present study contributes indeed to this discussion, by developing an agent-based model that reproduces the knowledge exchange process occurring during R\&D alliances.
Besides, we define here a novel indicator of network performance, based on the exploration of a knowledge space by the agents. In this way, our model is able to capture the existence of an optimal rate of alliance rewiring, as well as its dependence on the underlying technological regime.

\section{Model foundations}
\label{sec:model_foundations}
The microscopic rules of our agent-based model are inspired by a number of stylized facts, as well as theoretical speculations, in network evolution studies, opinion dynamics models, R\&D and collaboration networks. Below, we provide a brief description of every building block that we employ in the development of our model.

\paragraph{Monogamous network approximation.}
We model the formation of R\&D alliances between companies as a \textit{monogamous network}, i.e. a network in which every agent is linked to only one other agent at every time step \cite{vazquez2010epidemics,tessone2012synchronised}.
Inter-organizational networks are indeed proven to have low density, i.e. only a small fraction of all potential collaborations between companies are actually realized. The density of R\&D networks ranges from $0.1\%$ to $1\%$ for all industrial sectors, as shown in a previous study \cite{tomasello2013riseandfall}.
%Furthermore, the degree distributions of most empirical R\&D networks \cite{rosenkopf2007comparing,tomasello2013riseandfall} show that the vast majority of the nodes have only one partner, even in a time-aggregate representation.
However, some high-technology industries, such as Pharmaceuticals or Computers, although having low density, show high clustering and hierarchical structures, which are of fundamental importance for the dynamics of knowledge diffusion.
The ``hubs'' of these industrial sectors can actually have more than a hundred partners at the same time, with which they collaborate on different projects \cite{powell2005network,hanaki2010dynamics}.

Despite this empirical fact, we still use here the monogamous network as a modeling tool.
In order to have a more realistic picture, it should be noted that -- even though the agents have only one link at every time step -- they are allowed to change their partners in the following steps and can actually collaborate with many firms in a small time window.
Therefore, we propose as a possible extension to aggregate many network snapshots over time, similar to a previous theoretical study \cite{baum2010network} on R\&D networks; however, we deem this investigation to be beyond the scope of the present model and leave it for future research.

\paragraph{Position of companies.}
In the knowledge-based view of the firm, every company is endowed with a knowledge basis that uniquely identifies its resources and its capabilities. We assume that a firm is represented by an agent in our modeling framework, and associate it with a vector of $D$ components, each of which represents its share of knowledge in a given area. Furthermore, we directly associate these vectors to a metric \textit{knowledge space} in which the collaborations occur: every firm occupies a point in this $D-$dimensional space, whose coordinates are given by its knowledge vector.
Such an approach is similar to a more general model \cite{axelrod1997dissemination}, proposed in the broader context of social influence.
The concept of a metric knowledge space has already been used in one dimension \cite{groeber2009groups}, and in two dimensions \cite{fagiolo2003exploitation,baum2010network}. We generalize the dimensionality of the space to $D$.

The coordinates of every node can be thought of as the ratios of the corresponding firm's expertise along each of the $D$ dimensions of the space.
In order to have an empirical representation of these ratios, following an existing study \cite{tomasello2015effect_on_technology}, one can think of the different technological classes of which the International Patenting Classification (IPC) scheme is composed.
Just to give an example, the real IPC is divided into eight main categories, spanning from ``human necessities'' to ``electricity''.

Assuming that the classification scheme for our firms consists of $D'$ categories, then the $D$ values (with $D=D'-1$) would be the fractions of patents in each category to the total number of patents issued by the firm at hand.
It is important to note that such values are ratios, and not absolute measures of knowledge; therefore, there are no \textit{better} positions than others in the knowledge space that we utilize, but only \textit{different} positions, between which we can easily compute an appropriate measure of similarity.
These $D$ ratios are free to vary independently of each other in the interval $[0,1]$; the remaining $D'$-th, or ($D$+1)-th, knowledge component can be inferred from the main $D$ values through the bounding condition that the $D'$ values have to sum up to 1.

\paragraph{Alliance formation.}
In our monogamous network, all nodes are linked in pairs at every time step. We assume that two pairs of allied nodes mutually rewire their links at every time step with a given probability, and the new formed links are active if the Euclidean distance between the new partners is smaller than a threshold value.
Such a proximity condition models some existing theoretical arguments \cite{cohen89:_innov_learn,cohen1990absorptive}, highlighting that an interaction between two companies is profitable only if their \textit{absorptive capacity} is large enough or -- in other words -- their knowledge distance is small enough.
However, further studies \cite{lane1998relative,grant2004knowledge} have shown that there exists an inverted U-shaped relationship between the profitability of an alliance and the knowledge distance of two companies.
This means that, partners with a \textit{too small} knowledge distance (in other words, a \textit{too high} similarity) do not have any reason to establish an alliance.
Even though the selection strategy does not include such a curvilinear dependence, our agent-based model is able to capture this stylized fact, because we assume that the learning speed of the two agents (see Section \ref{sec:model} for more details) decreases with their knowledge distance, i.e. their learning potential.
Therefore, partners which are already similar in terms of knowledge bases do not significantly contribute to the increase of the relevant performance indicator in our model.

The choice of the Euclidean metric to compute this distance is quite realistic, even if it implies extensive information about the companies' mutual position in the knowledge space.
Indeed, obtaining detailed information about a company, its patent production, its scientific production and its activities in general is nowadays not only feasible -- thanks to the Internet -- but actually done by most firms willing to engage in an alliance \cite{ahuja2000collaboration,sampson2007diversity,baum2010network}.

The threshold value for the alliance establishment is supposed to model the technological regime that characterizes the collaboration network under examination.
A large interaction threshold means that the agents can establish active collaborations even with agents located far away in the knowledge space; this corresponds to an explicit knowledge regime, typical of a mature industry, where innovation is more routinized and mainly carried out by large incumbent firms, which have easy access to both similar and different firms (in terms of knowledge basis).
A small interaction threshold means that the agents can establish active collaborations only with agents located close in the knowledge space; this corresponds to a tacit knowledge regime, typical of a young industrial sector, where innovation is mostly carried out by small new-entrant firms, which have easy access to similar others in terms of knowledge basis.

\paragraph{Partner selection.}
The dynamics of alliance formation in the present model is assumed to be \textit{semi-random}, meaning that the rewiring of links between nodes occurs randomly and independently of the position of the nodes themselves in the knowledge space: we call this an \textit{exploration phase}. However, a link between two nodes is \textit{active} only if they are close enough in the knowledge space: if this happens, a so-called \textit{knowledge transfer phase} begins.
The rewiring mechanism does not intend to be a close representation of what happens in reality. It rather has the function of modeling the volatility of R\&D alliances, capturing the characteristic time scale at which firms decide to engage in a new alliance.
The second focal aspect that we want to model -- namely the formation of alliances at the right knowledge distance -- is instead fully captured by the threshold value for the potential partner's knowledge similarity.

\paragraph{Approaching in the knowledge space.}
Once a link has been established, we assume that a knowledge exchange between the partners takes place, causing their knowledge bases to become more similar and making them approach in the knowledge space.
This assumption is in line with the conceptualization of R\&D alliances as a means to exchange technological knowledge among firms \cite{mowery1998technological,owen2004knowledge,gomes2006alliances} and has already been used in a number of agent based models \cite{pyka2007,gilbert04:_agent_based_social_simulation_complexity,cowan2007bilateral}.
Besides, we argue that even when a knowledge transfer is not directly involved, for instance in a marketing or a cost-reducing alliance, an exchange of tacit knowledge can still take place between the two interested companies.

In our agent-based model, the speed at which the agents approach each other -- or, in other words, the rate at which they mutually learn form each other -- is governed by one parameter (further explanations follow below).
The closer the two agents are, the smaller their approaching speed becomes, as they are depleting the potential for mutual learning.
In addition, it should be noted that our work studies a scenario in which the knowledge spillovers occurring in a R\&D alliance cause the partners to exchange knowledge along every dimension, not limiting the knowledge transfer to a single area of expertise -- i.e.~one of the $D$ dimensions in the knowledge space.
Practically, this means that in every time step of our computer simulations the $D$ knowledge ratios of every pair of allied firms modify their values and become more similar; this approach is similar to a previous model \cite{baum2010network}.

\paragraph{Exploration of the knowledge space.}
Finally, we want to study the performance of the whole collaboration network as a function of the relevant model parameters. The indicator we propose to measure such a performance takes into account the global knowledge exploration of the system, i.e. it quantifies the distance traveled by all agents during the evolution of our simulated R\&D network.
In our model, we consider that the knowledge exploration itself is represented by the motion in the space, which is fully captured by this indicator.
The underlying assumption is that the exploration of as many locations as possible is beneficial for the R\&D collaboration network, in that it allows the agents to come in contact with many technological opportunities, potentially leading to more frequent innovations \cite{fagiolo2003exploitation}.
Testing our model by means of computer simulations, we find that the rewiring of links and the mutual knowledge exchanges over time eventually lead the whole system to a steady state through a peculiar dynamics. The model and its results are presented in detail in the next two Sections.

\section{The model}
\label{sec:model}

Starting from the evidence and the arguments presented in the previous Section, we now present the implementation of the agent-based model. We consider a network composed of $N$ nodes, each representing an agent -- in the particular case of R\&D networks, a firm -- performing collaboration activities in a knowledge space. The model is implemented by means of computer simulations, consisting of a sequence of discrete time steps of length $\mathrm{d}t$. The microscopic interaction rules are described below.

\subsection{Exploration phase}

Every node $i$ is located in a metric space (henceforth, the knowledge space); this point has coordinates $\mathbf{x}_{i}$, identified by a vector of $D$ real numbers ranging from 0 to 1. As already explained, the coordinates of every node can be thought of as the ratios of the corresponding firm's expertise along each of the $D$ dimensions of the knowledge space. At the initial stage of every simulation, all the nodes' positions are drawn from a uniform distribution in the interval $[0,1]$.
\begin{equation} \label{eq:position}
 \mathbf{x}_{i} \equiv (x_{i1},x_{i2}, \dots , x_{iD}) \qquad i=1, \dots , N .
\end{equation}

\begin{figure}[htbp]
\begin{center}
%\vspace{0.35cm}
\includegraphics[width=0.7\textwidth,angle=0]{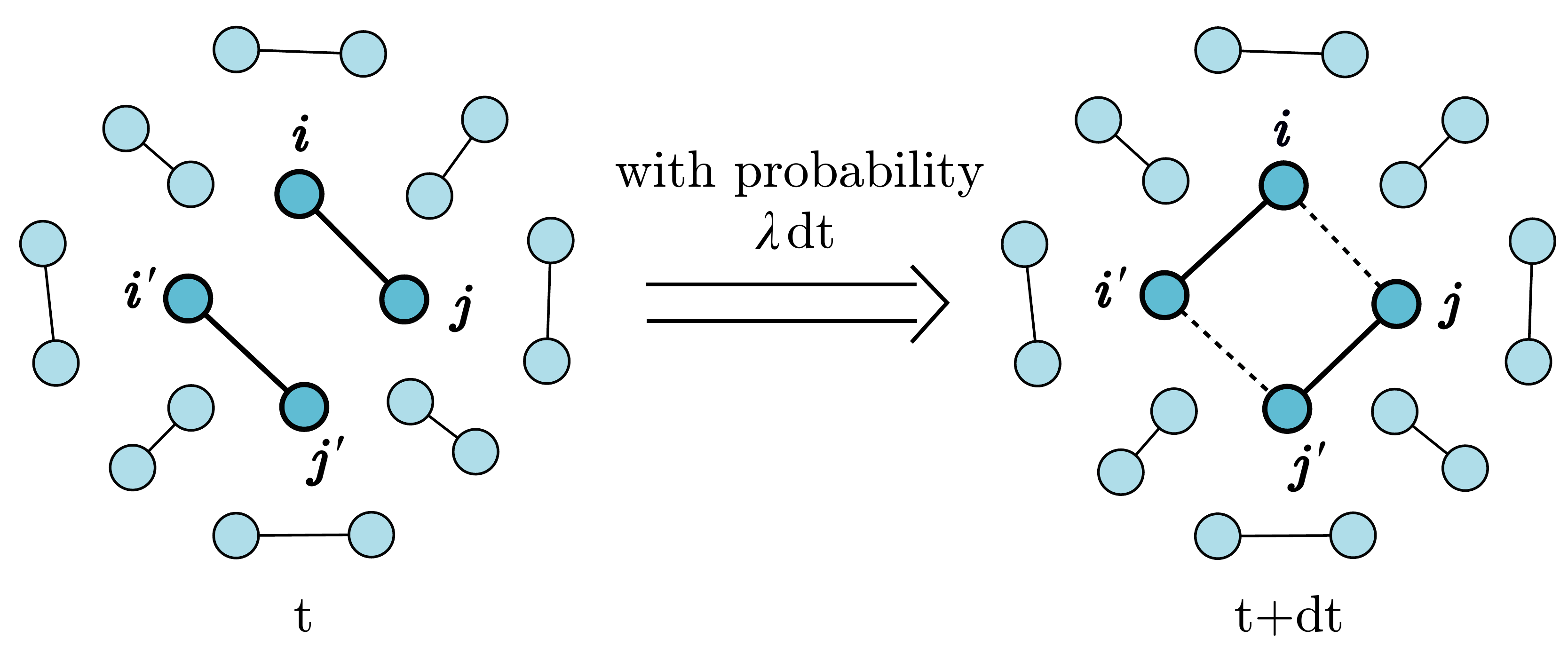}
%\vspace{-0.35cm}
\end{center}
\caption[Schematization of a link rewiring process.]{Schematization of a link rewiring between two pairs of connected nodes. At time $t$, the nodes $i$ and $j$ and the nodes $i'$ and $j'$ are linked in pairs. These two couples of nodes are selected and, with probability $\lambda\mathrm{d}t$, they switch links: at time $t+ \mathrm{d}t$ the nodes $i$ and $i'$ and the nodes $j$ and $j'$ are the new linked pairs. Obviously, with probability $1- \lambda \mathrm{d}t$, no rewiring happens.}
\label{fig:rewiring}
\end{figure}

All nodes in our R\&D network have the possibility to change their partner, thus generating a dynamic network topology. We model this by means of a link rewiring mechanism.
The time steps in our computer simulations have a duration equal to $\mathrm{d}t$; in each time step, two pairs of connected firms are randomly chosen and, with a rate $\lambda$, they rewire their links.
We call this process ``exploration phase'', and depict it in Fig. \ref{fig:rewiring}. Let us assume that the nodes $i$ and $j$ and the nodes $i'$ and $j'$ constitute the two linked pairs chosen at time $t$. With probability $\lambda \mathrm{d}t$, they mutually exchange their partners, and at time $t+ \mathrm{d}t$ the nodes $i$ and $i'$ and the nodes $j$ and $j'$ will form the new linked pairs. With probability $1- \lambda \mathrm{d}t$, instead, nothing happens and at time $t+ \mathrm{d}t$ the nodes $i$ and $j$ and the nodes $i'$ and $j'$ will still be respectively linked.

Such a random search for partners in the exploration phase has the function to model the volatility of R\&D alliances, capturing the characteristic time scale at which an agent decides to engage in a new collaboration.
The rate $\lambda$ can be indeed thought of as the inverse of the characteristic time elapsed before a firm takes part in a new alliance. Even though the potential partner is selected at random, the R\&D alliance will be actually ``active'' only if the partner fulfills a certain proximity condition in the knowledge space, as we will explain below.
Therefore, such an exploration is not fully arbitrary, and leads to the establishment of an actual collaboration only under specific conditions. It is worth mentioning that the results of our simulations remain qualitatively unchanged if we use any different random link creation process, as demonstrated in a previous study \cite{tessone2012synchronised}.

\subsection{Knowledge transfer phase}
The whole linking and rewiring process in our model occurs independently of the node knowledge positions, but their distance in the knowledge space has a determinant effect on the subsequent network dynamics.
Indeed, one of the key ingredients of our model is the existence of an optimal absorptive capacity for a profitable R\&D alliance between two firms. We assume that a link is \textit{active} if the corresponding pair exhibits a knowledge distance smaller than a given threshold value. If this proximity condition is not fulfilled, even though the corresponding nodes are connected, their link is considered to be \textit{inactive}, causing no effect at all on the system.
The proximity condition is evaluated for every pair of linked nodes $i$ and $i'$ as follows:
\begin{equation} \label{eq:interaction}
 |\mathbf{x}_{i}(t)-\mathbf{x}_{i'}(t)| < \epsilon \sqrt{D}
\end{equation}
where we employ the Euclidean distance $|\cdot|$, consistently with the assumption of evaluating the diversity of each firm's knowledge portfolio in all dimensions.
$\sqrt{D}$ is the maximum possible distance between two points in a $D$-dimensional Euclidean space. The parameter $\epsilon$, ranging from 0 to 1, is the threshold interaction radius inside which nodes are able to interact and collaborate profitably. Only links whose corresponding nodes fulfill this proximity condition are considered to be \textit{active}. Such an interaction radius can be associated with the knowledge regime characterizing the collaboration network under examination. A large $\epsilon$ means that the firms can potentially see and explore a large portion of the knowledge space, being the knowledge highly \textit{codified}. A small $\epsilon$ represents instead a regime of \textit{tacit} knowledge, where firms are able to establish alliances only if their technological positions are already close.

We assume that an R\&D alliance causes the two involved firms to pool their resources and their knowledge basis, thus approaching along every dimension in the knowledge space. Thanks to \textit{knowledge spillovers}, both firms will acquire common practices or a shared jargon, not limiting the knowledge transfer to that specific R\&D project that they have in common, as previously discussed.\footnote{However, we have also tested a scenario in which two allied firms exchange knowledge only in one dimension, thus moving in only one dimension of the knowledge space as well. The results remain qualitatively unchanged.}
If $i$ is an agent and $i'$ is its unique partner in the collaboration network at time $t$, both will move towards each other by identical paths in the knowledge space, provided that the proximity condition expressed in Eq. \ref{eq:interaction} holds. The model dynamics equation is the following: 
\begin{equation} \label{eq:knowledge_exchange_speed}
  \dot{\mathbf{x}_{i}}(t) = \mu \, [\mathbf{x}_{i'}(t)-\mathbf{x}_{i}(t)] , \qquad \mathrm{if} \quad |\mathbf{x}_{i'}(t)-\mathbf{x}_{i}(t)| < \epsilon \sqrt{D}
\end{equation}
where $\mu$ is defined as the \textit{learning rate} of the agents. This parameter is constant over time and for all nodes in the collaboration network, and can be thought of as the propensity of the agents to exchange knowledge with their partners, thus making their knowledge bases more similar over time.
It should be noted that the parameter $\mu$ is a rate, not a speed; the actual speed at which the corresponding nodes move in the knowledge space is given by the product of the rate $\mu$ and their distance: therefore, the farther they are in the knowledge space, the faster they approach. When their distance decreases, so does the potential for new learning from the collaboration, and the approaching speed drops consequently. This interpretation is clear in Eq. \ref{eq:approaching_global}, which represents the way we implement the model in computer simulations with discrete time steps of length $\mathrm{d}t$. The evolution of every agent's position $\mathbf{x}_{i}$ can be expressed as:
\begin{equation} \label{eq:approaching_global}
  \mathbf{x}_{i}(t+ \mathrm{d}t) = \mathbf{x}_{i}(t) + \mu \, \mathrm{d}t ~ [\mathbf{x}_{i'}(t)-\mathbf{x}_{i}(t)]
\end{equation}

\begin{figure}[h]
\begin{center}
%\vspace{0.35cm}
\includegraphics[width=0.6\textwidth,angle=0]{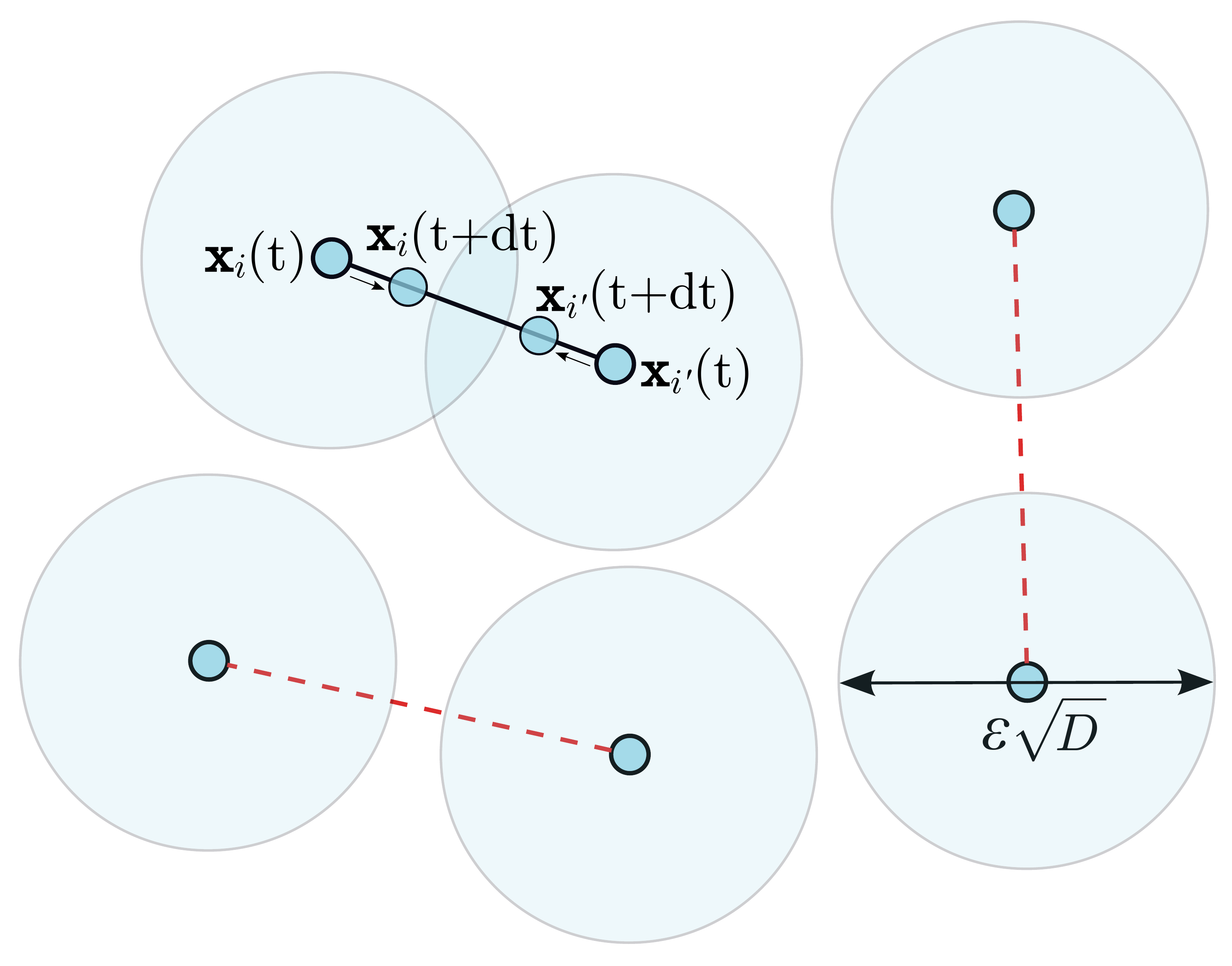}
%\vspace{-0.35cm}
\end{center}
\caption[Schematization of the knowledge exchange process in a bi-dimensional knowledge space.]{Schematization of the knowledge exchange process in a bi-dimensional space ($D=2$). At time $t$, the agents $i$ and $i'$ are linked and their distance $|\mathbf{x}_{i'}(t)-\mathbf{x}_{i}(t)|$ is smaller than $\epsilon \sqrt{D}$; consequently, at time $t+ \mathrm{d}t$, their positions $\mathbf{x}_{i}(t+ \mathrm{d}t)$ and $\mathbf{x}_{i'}(t+ \mathrm{d}t)$ will be closer in the knowledge space. The picture includes other pairs of connected agents, whose distance is larger than $\epsilon \sqrt{D}$. Therefore, these links are \textit{inactive} (depicted in dashed red lines) and do not originate any motion in the knowledge space.}
\label{fig:knowledge_exchange_approach}
\end{figure}

We depict such knowledge exchange mechanism in Fig. \ref{fig:knowledge_exchange_approach}. The nomenclature and the meaning of all the model parameters we introduced in this Section are summarized in Table \ref{table:knowledge_exchange_parameters}.

\setlength{\tabcolsep}{8pt}
\renewcommand{\arraystretch}{1.1}

\begin{table}[h]
\footnotesize
\centering
\begin{tabular*}{0.98\linewidth}{ c | l | l }
\hline

\textbf{Parameter} & \textbf{Meaning} & \textbf{Type of parameter} \\
\hline

$N$ & Number of agents (system size) & Static \\
$D$ & Dimensionality of the metric knowledge space & Static \\
$\epsilon$ & Agents' interaction radius (knowledge regime) & Static \\
$\lambda$ & Link rewiring rate & Network dynamics \\
$\mu$ & Approaching rate in the knowledge space & Network dynamics \\
%[0.1ex]
\hline
\end{tabular*}

\caption[Model parameters and their description.]{Model parameters and their description. The ``static'' parameters are associated with the system structural features, while the ``network dynamics'' parameters define the characteristic speed at which the system evolves.}
\label{table:knowledge_exchange_parameters}
\end{table}

\section{Results}

In order to test our model, we have performed extensive computer simulations by applying the dynamics presented in Section \ref{sec:model} and varying the values of the relevant parameters. In particular, we vary the size $N$ of the network from $10$ to 2,000 nodes, the dimensionality $D$ of the knowledge space from $1$ to $50$, the interaction threshold radius $\epsilon$ from $0$ to $1$, the learning rate $\mu$ from $10^{-3}$ to $10^3$ and the rewiring rate $\lambda$ from $10^{-3}$ to $10^4$.
All of these parameters are explored in discrete intervals, whose width is appropriately chosen -- as we discuss below in more detail.
The value of $\mathrm{d} t$ is instead fixed for all our computer simulations to a value of 0.0001. All values are expressed in arbitrary units.

\paragraph{Main model parameters and their meaning.}
We argue that the network evolution is essentially characterized by two driving forces with overall opposite effects. The first one is the formation of active links (i.e. the establishment of profitable alliances or collaborations); this force tends to push agents closer in the knowledge space, given the resulting approaching motion.
The second force is the link rewiring (representing the dissolution of old collaborations and the formation of new ones), that stimulates the agents to explore new portions of the knowledge space. This force could result in an faster overlap of every agent's knowledge position, but it could also result -- under certain conditions -- in preventing the agents from converging to a knowledge attractor, thus keeping them far-between in the knowledge space.

These competing forces are associated with the two model dynamics parameters, respectively the approaching rate $\mu$ and the link rewiring rate $\lambda$. However, it is clear that the relation between these two parameters will substantially affect the emergent properties of the system. What truly affects the resulting dynamics of the network are not the absolute values of the two rates $\mu$ and $\lambda$, but the ratio of the two.
Indeed, using a configuration with the same $\mu$ to $\lambda$ ratio, but with smaller absolute values, will only lead to a longer computer simulation (i.e. more discrete time steps are needed), without qualitatively changing the results. Therefore, in the continuation of the current study we present our findings by keeping the value of the learning rate fixed to $\mu=1$, and studying the effect of the dynamics parameter $\lambda$ only.

The second relevant model parameter on which we focus our attention is the threshold interaction radius $\epsilon$, a static parameter representing the knowledge regime in which the collaborating agents move. We explore a series of values ranging from a totally tacit knowledge regime ($\epsilon=0$) to a totally explicit one ($\epsilon=1$).

The effect of the dimensionality parameter $D$ and the network size $N$ is that of changing the characteristic density of the system, i.e.~the number of agents that can be found within the given interaction radius $\epsilon$.
Varying their values causes shifts in the trends of the relevant measures that we investigate (network performance, number of knowledge clusters and convergence time), without qualitatively affecting the results.
Indeed, the numeric values of the parameters that we present here do not have a specific meaning, and the ranges that we explore in the present study are only aimed at a clear and effective visualization of our findings.
For the sake of extendability and future empirical work, all our model parameters have a straightforward and natural match with real quantities, including the parameters whose dependencies are not directly studied here, such as the size $N$ of the system, its dimensionality $D$, or the learning rate $\mu$.
Should these parameters be matched with their empirical counterparts, all their values would assume a real, directly intelligible meaning.

One last remark has to be made about the number of agents $N$ in the system. In empirical collaboration networks, such a number is obviously dynamic, and not static.
However, the incorporation of a dynamic system size in our agent-based model would deeply modify it, and shift the focus away from the investigation of the agents' knowledge exploration.
Such an extension could indeed constitute a second, distinct study, and we leave it for future research.\footnote{
Indeed, there exists a work \cite{tomasello2015effect_on_technology} which attempts to study a similar issue of knowledge exchange in a system where the time scales of the agents' interactions and their entry/exit in the network is not decoupled; a dynamic number of agents can partially be captured by that model through a quantity called ``activity''. Such a study could be extended and improved with the findings deriving from the present model.
}

In the continuation of the present study, we select a network composed of $N=200$ nodes and a knowledge space with $D=10$ dimensions, to present our results in the most effective way possible.

\paragraph{Network performance.}
The variable that we investigate as indicator of the network performance is the \textit{mean knowledge path} $\mean{K}$ of the collaborating agents. We define the path covered by every agent in the knowledge space $K_{i}$ as the sum of all the distances that the agent travels in every time step of the simulation:
\begin{equation} \label{eq:knowledge_path}
 K_{i} = \int_{t=0}^{T_{\mathrm{max}}} \abs{\dot{\mathbf{x}_{i}}(t)} \, \mathrm{d}t
\end{equation}
where $T_{\mathrm{max}}$ is the duration of an entire computer simulation. It should be noted that the measure $\abs{\dot{\mathbf{x}_{i}}(t)}\, \mathrm{d}t$ is a positive scalar and expresses the actual distance traveled by the agent $i$, differently from its net displacement $\dot{\mathbf{x}_{i}}(t)\, \mathrm{d}t$, which is a vectorial quantity. The measure $K_{i}$ is then averaged over all the $N$ network agents to obtain the mean knowledge path $\mean{K}=N^{-1} \cdot \sum_i{K_i}$.
We hypothesize that this measure can provide a meaningful indication of the macroscopic system performance, because -- as already discussed in Section \ref{sec:model_foundations} with respect to the microscopic level -- firms are proven to innovate more when they come in contact with more technological opportunities. Therefore, we assume that a higher value of $\mean{K}$, i.e. a higher distance explored in the knowledge space, corresponds to a higher network performance. We argue that the same reasoning can be as well extended to other types of collaborations that involve learning and/or knowledge exchange processes.

\begin{figure}[h]
\begin{center}
%\vspace{0.35cm}
\includegraphics[width=0.99\textwidth,angle=0]{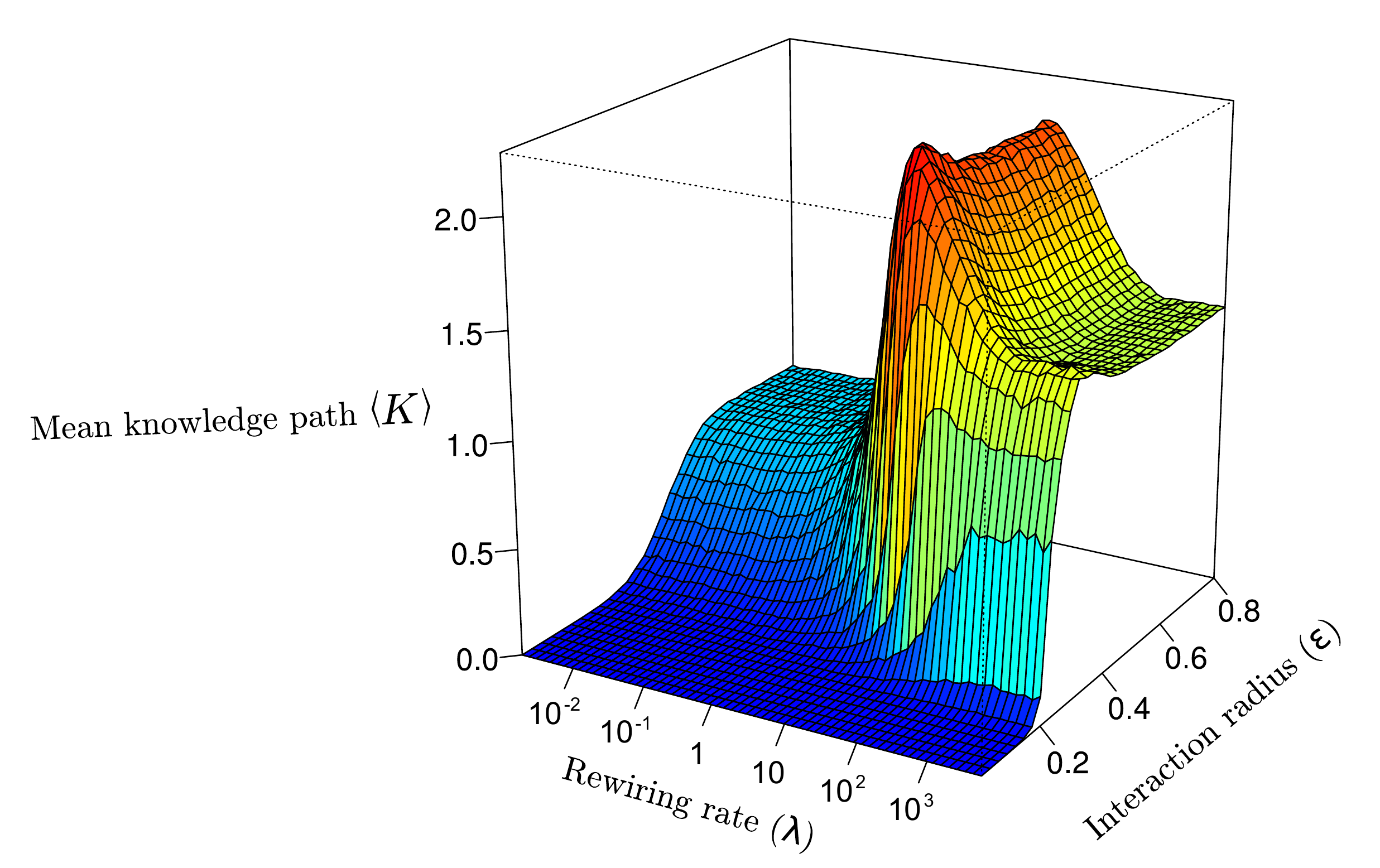}
%\vspace{-0.35cm}
\end{center}
\caption[Mean knowledge path as a function of the link rewiring rate and the agent interaction radius.]{Mean knowledge path $\langle K \rangle$ (displayed by means of both the z-elevation and the color scale), as a function of the rewiring rate $\lambda$ and the interaction radius $\epsilon$. The R\&D network under examination has $N=200$ nodes and learning rate $\mu=1$, in a $10-$dimensional knowledge space. We generate 1000 simulations for each parameter set and then average the results.}
\label{fig:knowledge_path_results}
\end{figure}

We present the results in Fig. \ref{fig:knowledge_path_results}, for a representative network of $N=200$ agents moving in a knowledge space with $D=10$ dimensions. As already mentioned, the parameter $\mu$ is fixed to 1, and we study the dependence of $\mean{K}$ on the dynamics parameter $\lambda$ and the static parameter $\epsilon$.
%For a two-dimensional representation of the same results, see Appendix \ref{appendix5}.

We find that the mean knowledge path $\langle K \rangle$ exhibits a peak in correspondence of specific values for both the rewiring rate $\lambda$ and the interaction radius $\epsilon$. In the case that we present in Fig. \ref{fig:knowledge_path_results}, these values are $\lambda \simeq 10$ and $\epsilon=0.25$, located in the red area of the plot.

Taking a closer look at the network performance, we find that $\langle K \rangle$ shows a monotonic growing trend as a function of $\lambda$, when the interaction radius $\epsilon$ is lower than a certain value $\epsilon^{*}$ (in our example, $\epsilon^{*} \simeq 0.25$). When fixing the interaction radius to larger values $\epsilon \ge \epsilon^{*}$, we do instead find that $\langle K \rangle$ exhibits a peak as a function of $\lambda$.
This means that, as the knowledge regime becomes more explicit, and the agents are allowed to form active collaborations even with distant partners in terms of knowledge basis, there exists a rewiring rate maximizing the distance actually explored by the agents in the knowledge space.

The behavior of the mean knowledge path $\langle K \rangle$ can also be described as a function of the interaction radius $\epsilon$, while keeping the rewiring rate $\lambda$ fixed. We find that $\langle K \rangle$ grows with $\epsilon$ to a saturation level (when $\epsilon > 0.5$), if the rewiring rate is small ($\lambda < 1$, for the case under study).
If we fix the rewiring rate $\lambda$ to a value larger than 1, we find instead that $\mean{K}$ increases to a peak, in correspondence to $\epsilon \simeq 0.35$, and then decreases again to stabilize for $\epsilon > 0.5$.
This means that, when the characteristic alliance rewiring rate of the network is greater than the characteristic learning rate of the agents -- i.e. when the search for new alliance partners is predominant over the learning mechanism -- there exists a specific threshold interaction radius (corresponding to a moderately explicit knowledge regime) maximizing the distance covered by the agents in the knowledge space.

This finding can be interpreted in terms of emergence of knowledge attractors in the system (please refer to the paragraph below for more detailed explanations).
In a regime dominated by the exploratory search for new alliance partners ($\lambda > 1$), increasing the interaction radius $\epsilon$ is beneficial for the system, because the agents can form alliances with more distant partners, and therefore travel longer distances in the knowledge space.
However, a too high interaction radius -- together with $\lambda$ values greater than 1 -- causes the emergence of just one central attractor, toward which the agents converge quickly, without traveling too much distance in the knowledge space.
This happens because the agents have the chance to interact with all others, thus quickly uniforming their knowledge bases.

On the other hand, when looking at the rewiring rate rather than the interaction radius, the explanation of the inverted U-shaped network performance is dynamical instead of spatial.
Here, a high rewiring rate is beneficial because it brings the agents in rapid contact with other agents situated far away in the knowledge space, increasing -- on average -- the distance that they travel.
However, a too high rewiring rate (combined with an intermediate value of $\epsilon$) results in the emergence of several attractors in distinct locations of the knowledge space, as we report in detail in the next paragraph.
This causes each of the agents to converge to one of the many attractors, rather than the unique central attractor, which would instead emerge in case of a lower rewiring rate.
Eventually, this translates into a lower distance traveled by the agents, if compared to a medium-rewiring-rate scenario, where all agents would globally travel a longer distance to reach the central attractor.

Whether the distance traveled in the knowledge space is a better performance indicator than the number of emerging knowledge attractors is still an open question, which probably requires a case-by-case discussion depending on the system under examination.
In any case, the analysis of the mean knowledge path in the system cannot be decoupled from the analysis of the knowledge attractors emerging in the system.
In the next paragraph, we examine in more detail how our agent-based model can explain their formation and evolution.

\paragraph{Knowledge clusters and attractors.}
We here investigate a second emerging property of the system, namely the \textit{number of knowledge clusters} appearing in the network at the end of every model run.
We define a knowledge cluster as a group of nodes whose mutual distances are smaller than $\epsilon$. Moreover, the distance between every node in that cluster and every node outside of that cluster has to be larger than $\epsilon$, meaning that all the agents in the cluster will asymptotically converge to one point in the knowledge space, and no further inclusion of any other agent in the cluster is possible.
We call such a point a knowledge attractor, or simply an attractor.

\begin{figure}[h!]
\begin{center}
%\vspace{0.35cm}
\includegraphics[width=0.6\textwidth,angle=0]{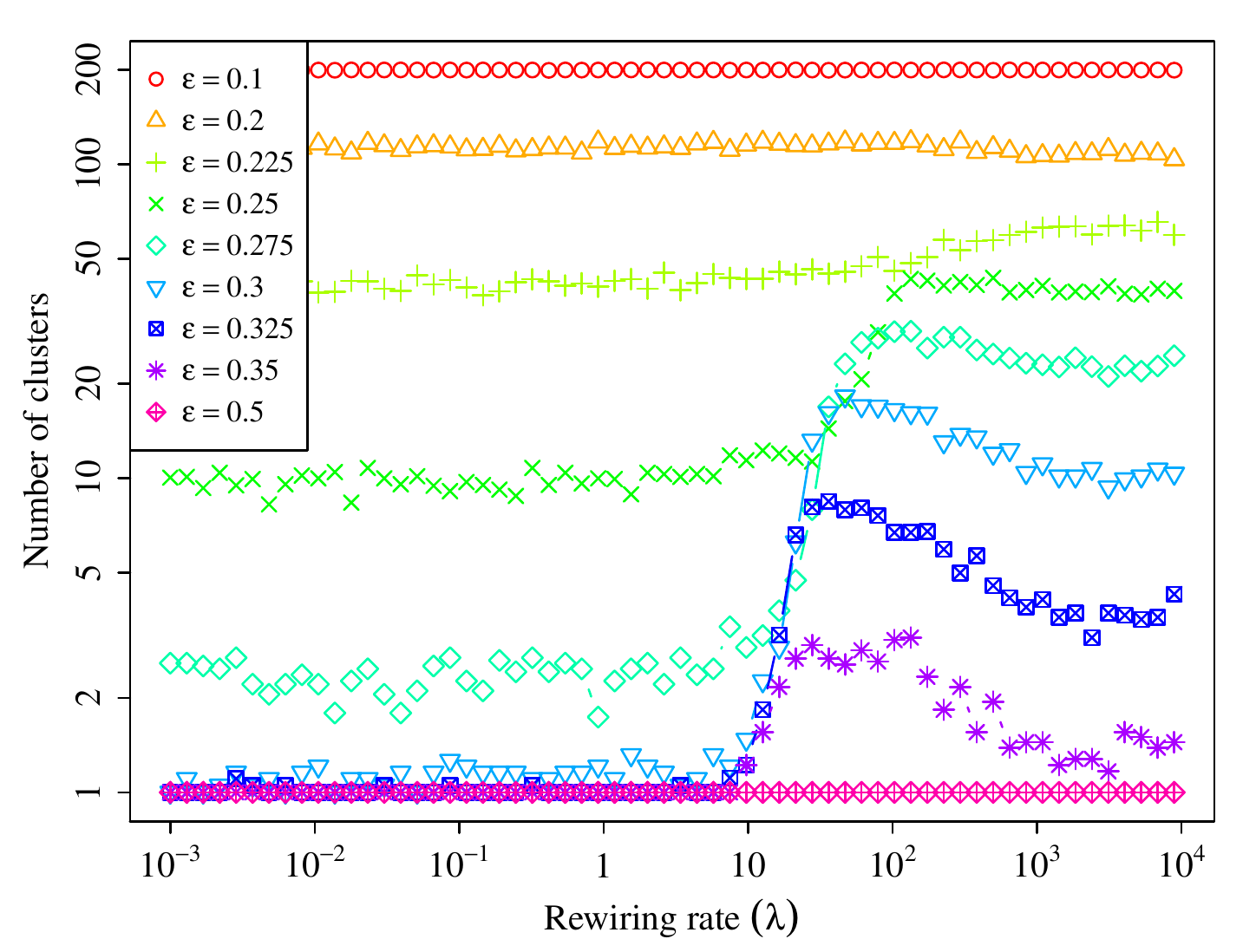}
%\vspace{-0.35cm}
\end{center}
\caption[Number of knowledge clusters.]{Number of knowledge clusters as a function of the rewiring rate $\lambda$, for a set of representative values of the interaction radius $\epsilon$. The network under examination has $N=200$ nodes and learning rate $\mu=1$, in a $10-$dimensional knowledge space. We generate 1000 simulations for each parameter set and then average the results.}
\label{fig:knowledge_clusters_results}
\end{figure}

It is clear that the maximum possible value of knowledge clusters equals the number of nodes $N$; we expect to observe such a value in correspondence with a low value of the interaction radius $\epsilon$, when the agents are virtually unable to establish active links.
Likewise, the minimum possible number of knowledge clusters equals 1; we expect to observe such a value in correspondence with high values for the interaction radius $\epsilon$, when most established collaborations are active, thus facilitating the convergence of all agents toward one knowledge attractor.
Similarly to the mean knowledge path, we present our results in Fig. \ref{fig:knowledge_clusters_results}, for a network of $N=200$ agents in a knowledge space with $D=10$ dimensions; $\mu$ is fixed to $1$.

We find that the number of clusters generally increases by decreasing the interaction radius $\epsilon$. As expected, an extreme case occurs for $\epsilon=0$ (completely tacit knowledge regime, where any interaction is by definition impossible), in which we have as many clusters as agents -- independently of the rewiring rate $\lambda$. The other extreme case occurs for $\epsilon \ge 0.5$ (highly explicit knowledge regime), in which all the nodes interact between each other converging in only one cluster -- again, independently of $\lambda$.

\begin{figure}[h!]
\begin{center}
%\vspace{0.35cm}
(a)\includegraphics[width=0.45\textwidth,trim=50 80 40 60]{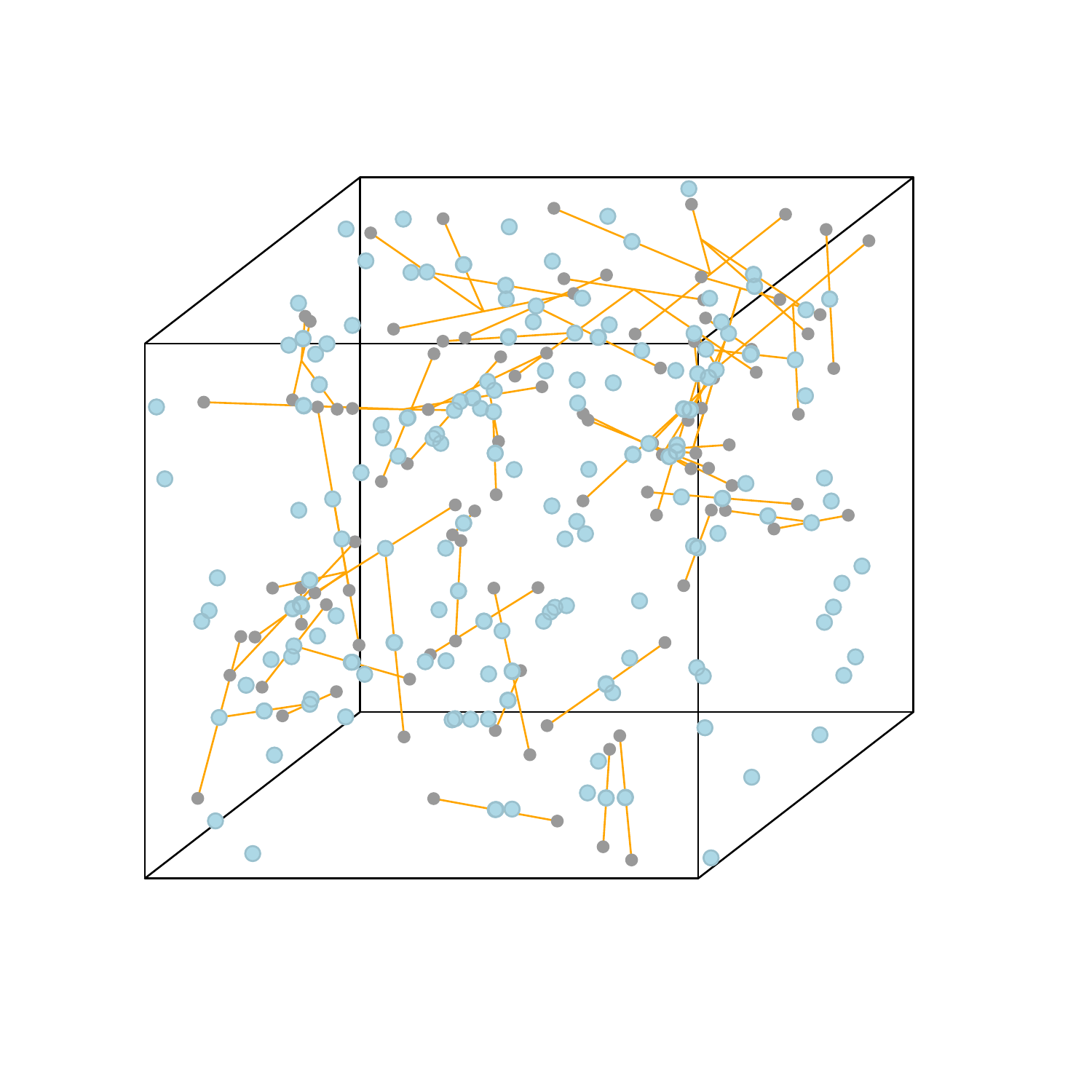}
(b)\includegraphics[width=0.45\textwidth,trim=50 80 40 60]{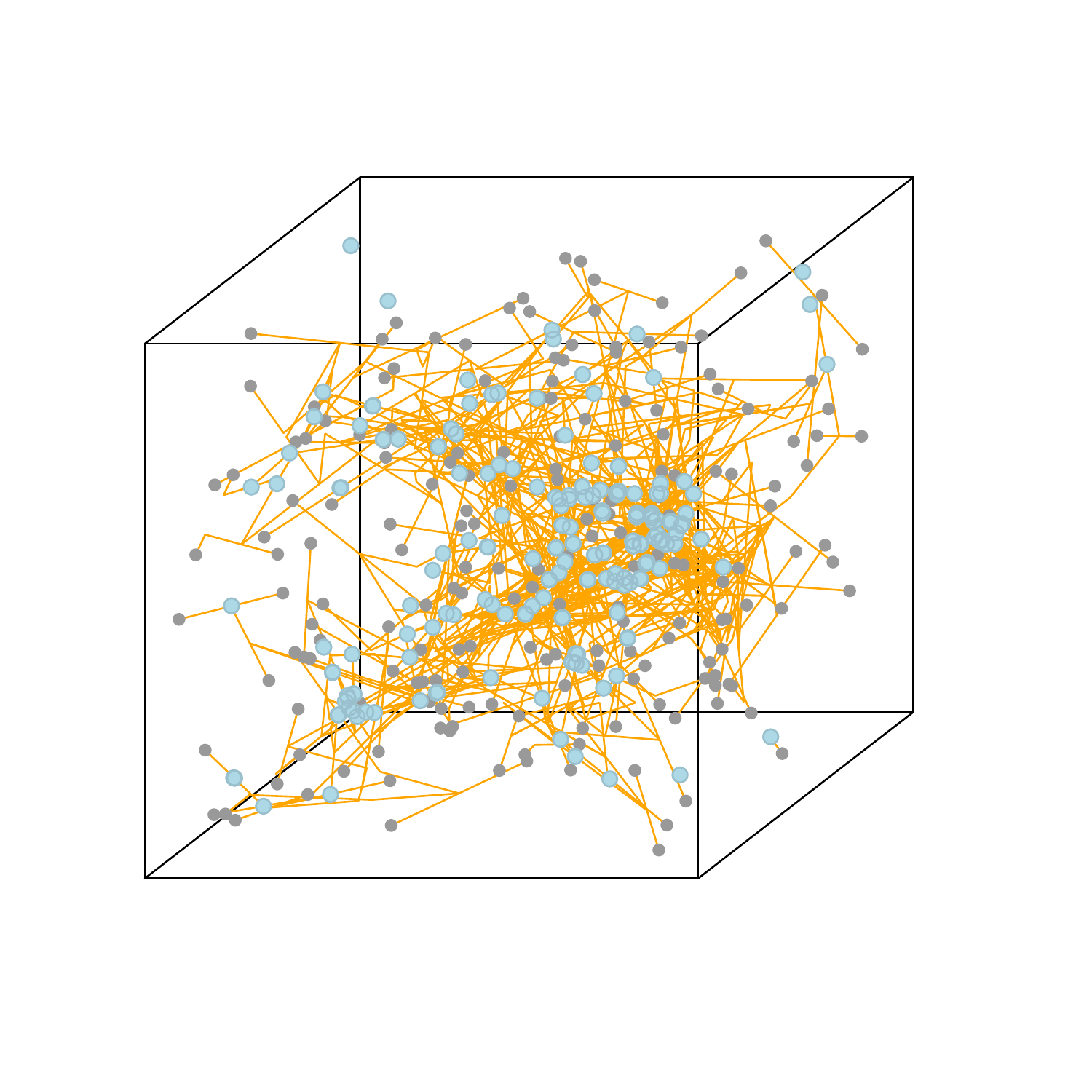}
(c)\includegraphics[width=0.45\textwidth,trim=50 80 40 50]{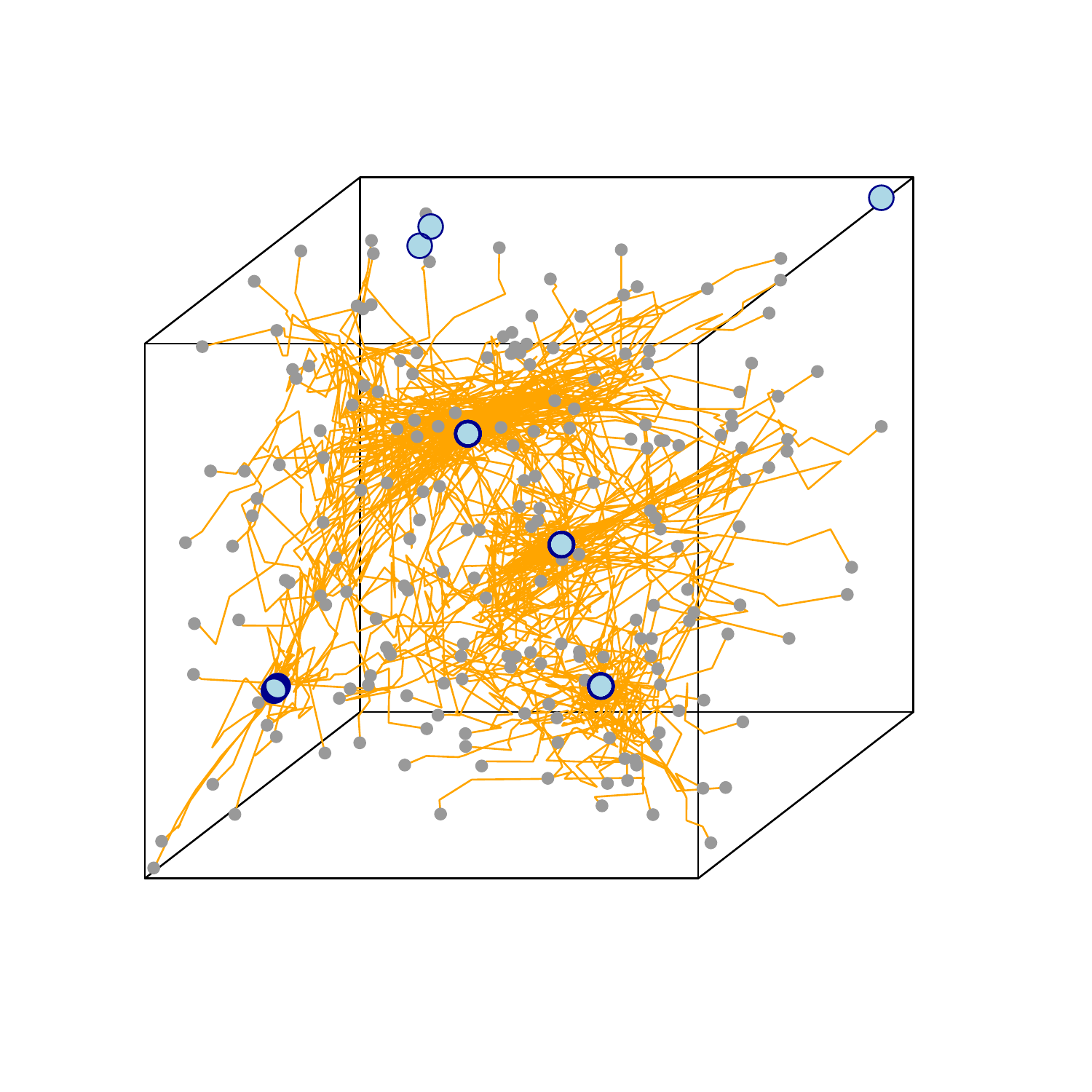}
(d)\includegraphics[width=0.45\textwidth,trim=50 80 40 50]{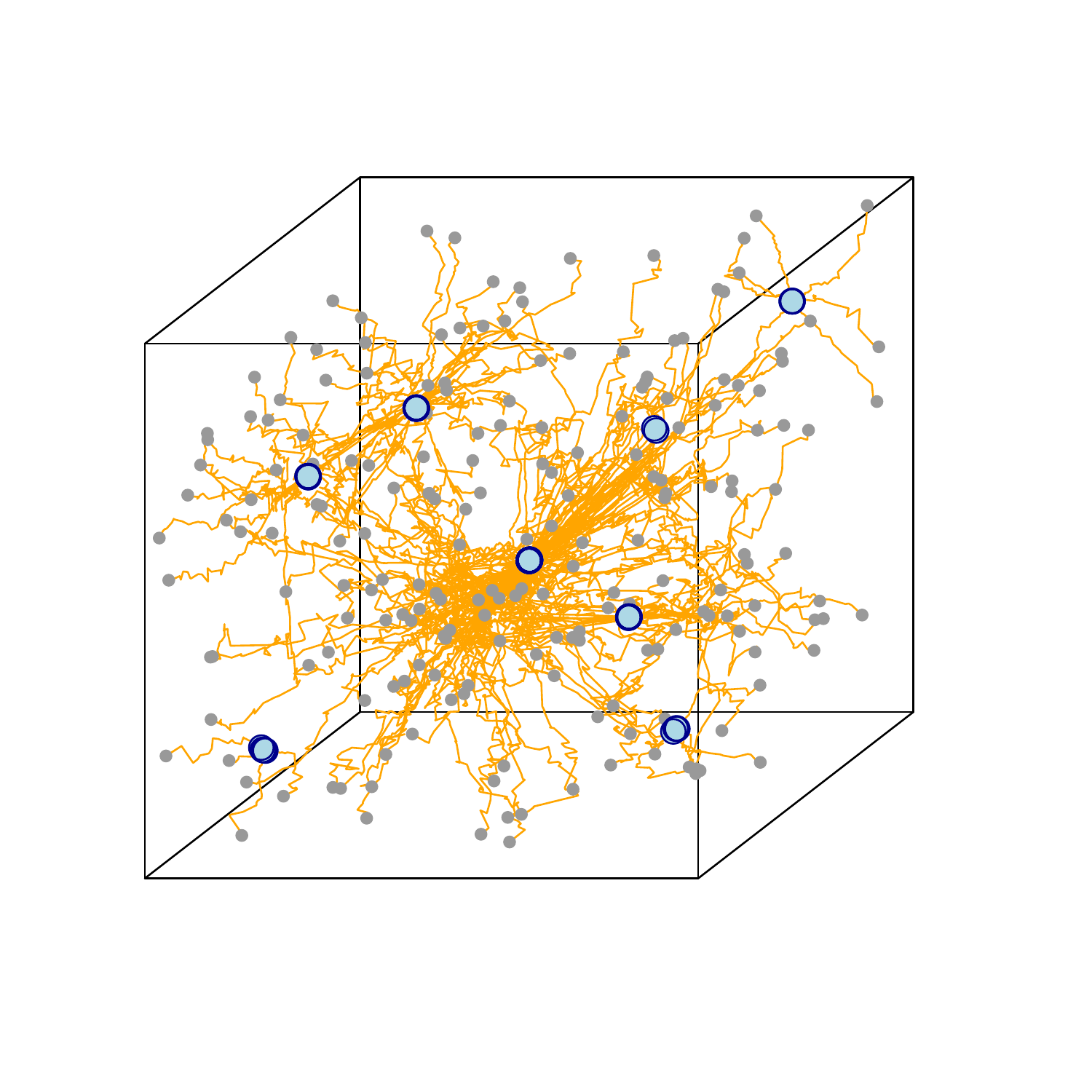}
%\vspace{-0.35cm}
\end{center}
\caption[Knowledge trajectories for a network with $N=200$ nodes in a 3-dimensional knowledge space.]{Knowledge trajectories for a network with $N=200$ nodes and learning rate $\mu=1$. For the sake of visualization, here we use a knowledge space with $D=3$ dimensions, easily representable as a cube. The initial positions of the nodes are depicted with gray dots, their trajectories with orange lines, and their final positions with blue dots. If the final position of an agents corresponds to a knowledge attractor, we depict this with a circled, dark blue dot. We keep the threshold interaction radius constant to $\epsilon = 0.2$, and show four cases corresponding to rewiring rate $\lambda$ equal to: (a) $1$, (b) $10$, (c) $10^2$, (d) $10^3$. In the cases (a) and (b), the agents are ``frozen'' and dispersed in the knowledge space, and they all belong to a unique, giant knowledge cluster. In the cases (c) and (d), the faster rewiring rate allows for the emergence of several distinct knowledge attractors, to which 
the agents converge through longer, meandering trajectories.}
\label{fig:dendrites}
\end{figure}

Noteworthy, for intermediate values of $\epsilon$, we observe an interesting dependence of the number of knowledge clusters on the rewiring rate $\lambda$. When $\lambda$ is low (i.e. comparable in magnitude with $\mu$), we find the existence of one or very few knowledge clusters, because the overall effect of such a slow rewiring rate is that all nodes tend to get closer in the knowledge space before the corresponding links are cut and rewired.
As a result, all nodes are eventually part of the same knowledge cluster. From the visual examples in Fig. \ref{fig:dendrites} (a) and (b), we can observe that such clusters are dispersed in the knowledge space, and the presence of a central attractor is not visually detectable, even though all the agents are in principle within interaction distance.
What happens, in fact, is that every pair of agents converges to the midpoint of the segment connecting them; the system then ``freezes'' in this configuration, being the rewiring rate too low to allow for new collaborations and new explorations within a meaningful time frame.

It should be noted that, if the computer simulations were allowed to last longer, the agents could in principle converge to the central attractor above mentioned, thus making it better visible in the knowledge space.
However, this would result only in a different visual outcome and would change neither the number of detected knowledge clusters (all the agents are included in one giant cluster anyway; see definition above), nor the way the agents travel in the space (i.e. converging to the midpoint of the connecting segments and then freezing there for most of the time), thus not affecting our findings about the network performance.

When the value of $\lambda$ increases, instead, we observe the formation of a higher number of knowledge clusters. These clusters are well delimited in the knowledge space and, as we show in the examples of Fig. \ref{fig:dendrites} (c) and (d), the presence of attractors is visually evident.
Such a non-trivial effect derives from the fact that the nodes cut their links and form new ones before the approaching mechanism with the previous partner is complete, thus traveling with a peculiar meandering trajectory, also clearly visible in Fig. \ref{fig:dendrites} (c) and (d).

%Interestingly, the effect of experiencing more alliances with different partners is therefore the emergence of distinct knowledge attractors, rather than causing all firms to converge towards the same knowledge attractor, thus uniforming their knowledge bases.

It is interesting to note how such peculiar trajectories result both in a longer traveled distance in the knowledge space and in the emergence of several attractors, occupying different regions of the space.
However, when $\lambda$ increases above a given threshold -- in accordance with our findings on the network performance -- the longer meandering trajectories are no longer able to compensate for the shorter distances globally traveled by the agents.
The emergence of several distinct attractors, indeed, causes the agents to travel a shorter distance, on average, before converging to one of them.
For the case examined in Fig. \ref{fig:knowledge_path_results}, this results in a decreasing network performance when the rewiring rate is higher than $\sim 10$, interestingly the value in correspondence of which the number of clusters explodes (see Fig. \ref{fig:knowledge_clusters_results}).

\paragraph{Convergence time.}
We find that the network dynamics generated by the model eventually converges to a steady state, in which all the agents occupy one or more fixed positions, and no further collaborations, nor motion in the knowledge space, are possible. In other words, such a steady state represents a configuration in which the collaborating agents have depleted all the potential for new knowledge exchange.

\begin{figure}[h!]
\begin{center}
%\vspace{0.35cm}
\includegraphics[width=0.6\textwidth,angle=0]{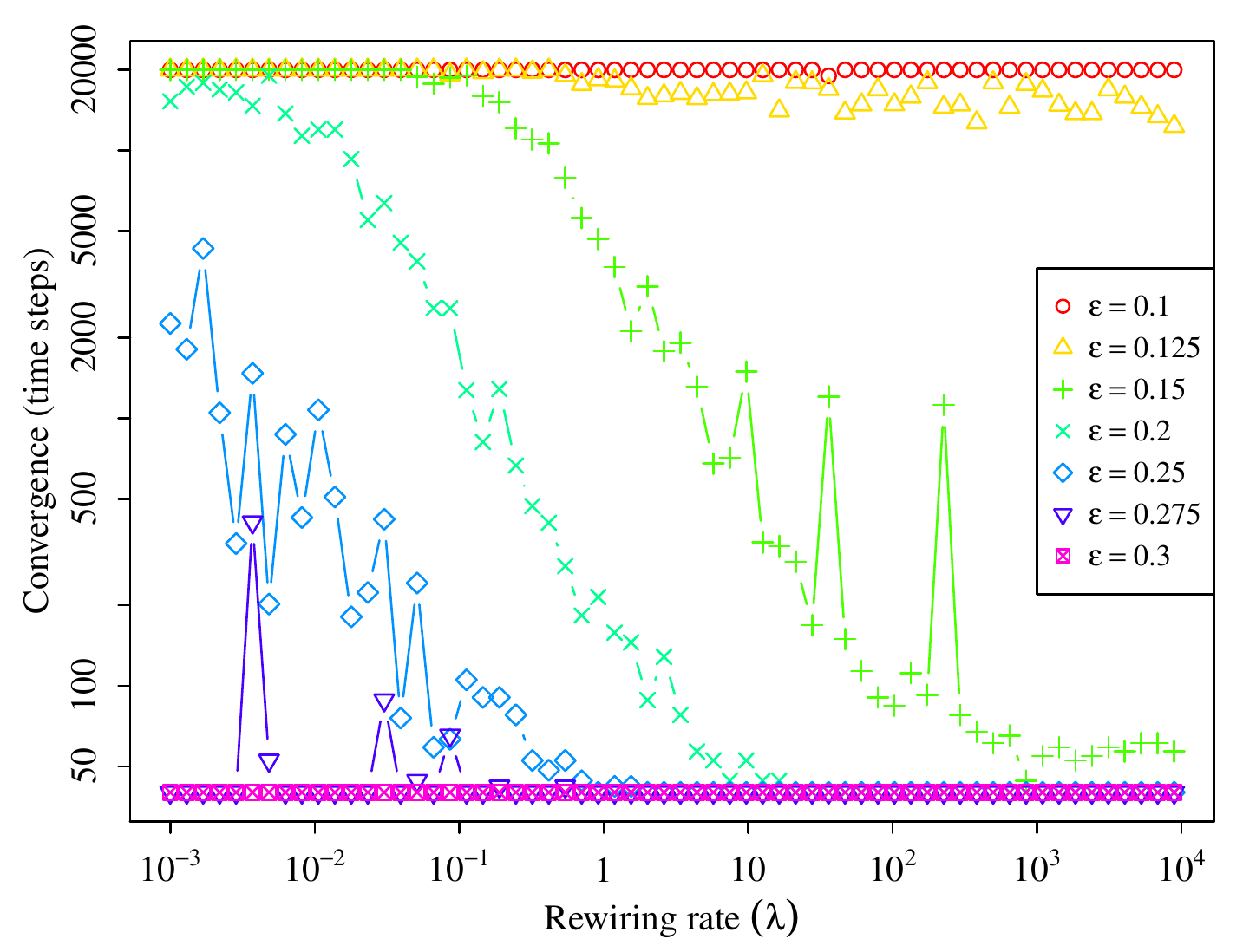}
%\vspace{-0.35cm}
\end{center}
\caption[Convergence time of the system.]{Convergence time as a function of the rewiring rate $\lambda$, for a set of representative values of the interaction radius $\epsilon$. The network under examination has $N=200$ nodes and learning rate $\mu=1$, in a $10-$dimensional knowledge space. We generate 1000 simulations for each parameter set and then average the results.}
\label{fig:knowledge_convergence_time_results}
\end{figure}

We define a convergence criterion based on the agents' motion in the knowledge space, and assume that the steady state is reached if the total knowledge path traveled by all the agents in the last time step is smaller than the 0.5\% of the knowledge path covered by the agents in the last 500 time steps.
Indeed, all of the network measures described above are computed only after the steady state is reached. In Fig. \ref{fig:knowledge_convergence_time_results}, we show the trend of the convergence time as a function of $\lambda$ and $\epsilon$, for the same representative network that we have studied before.

On the one hand, we find that all the relevant parameter configurations reach a steady state before the computer simulation ends. Indeed, the parameter combinations that are not able to reach a steady state before the end of the simulation (those with $\epsilon < 0.15$ or generally low $\lambda$) are the ones generating the lowest values of mean knowledge path, for the reasons we previously discussed. Therefore, we forcedly stop all computer simulations after 20,000 time steps, affecting only a small fraction of the parameter space and not influencing our results.
It should be noted that we aim at developing a model that has the potential to be validated against real data.
%a value of lambda as high as $10^4$ (corresponding to one rewiring per time step, ) would not make empirical sense, and only lambda values up to $\sim 10^2$ would be realistic. A solution would be to use a different time mapping, or to exclude the parameters values that do not make empirical sense.
Because of this, we are interested in configurations which can be considered as a pseudo steady state within a meaningful time frame; this means a few orders of magnitudes ($\sim 4$, in our case) longer than the characteristic interaction time.
This is consistent with most of the empirical datasets on R\&D or collaboration networks, whose typical observation length (a few decades) is around 4 orders of magnitude larger than their granularity (1 day).\footnote{
In our model, the quantity $\lambda \cdot \mathrm{d}t$ represents, at its core, a simplification of a Poisson process.
In a hypothetical future work, when validating the model on empirical data, one should pay careful attention to the utilized numeric values. For instance, if one assumes that a simulated time step equals one day in reality, the values of both $\mathrm{d}t$ and $\lambda$ have to be chosen in such a way that the number of simulated rewirings is comparable with the number of empirically observed inter-firm alliances in the given observation period, to allow for a matching of the granularity of the events.
}
Therefore, when true convergence to a stationary state requires letting the system evolve for a much higher number of time steps, this could be deemed unrealistic. 
%However, this leaves our results unchanged.

On the other hand, we find an unexpected trend of the convergence time as a function of $\lambda$ for some parameters combinations. One would expect that the convergence time decreases proportionally to $1/\lambda$, being the inverse of the rewiring rate a measure of the characteristic time of the system for a complete interaction between all agents.
More precisely, considering that $\lambda$ is the characteristic rate for one interaction between two agents in the system, its inverse $1/\lambda$ is the time needed, on average, by each agent to interact with all other agents in the system.
This means that, if one neglects all the network effects and the complex interdependencies of the model, the characteristic convergence time of the system would be proportional to $1/\lambda$.

However, we observe such a trend only for the extreme cases of highly explicit knowledge regimes, corresponding to $\epsilon \ge 0.5$, where the network effects are minimum and a complete interaction between all agents does indeed take place, because of the very large interaction radius.
Instead, we find a peculiar trend of the convergence time as a function of $\lambda$ for all the other values of $\epsilon$, showing plateaux for high values of $\lambda$. This means that the complex network dynamics, in the presence of certain approaching and link rewiring rates, can slow down the convergence of the system; as a result, the steady state is reached later than the characteristic time $1/\lambda$ would suggest.

\section{Discussion}

In this study, we have developed an agent based model of knowledge exchange and dynamic rewiring of R\&D alliances. Our novel contribution has been to explicitly model how agents move toward each other in a metric knowledge space. In addition, we have studied the co-evolution and the interdependencies of such a process with a dynamically evolving network structure.

By studying the interactions of a set of agents in a metric knowledge space via computer simulations, we have found that the system follows a peculiar dynamics and reaches a steady state in which the agents cluster around a set of emerging attractors.
The model parameters that determine the overall properties of the system are the link rewiring rate of the network and the agents' interaction radius.

\paragraph{Our findings.}
We have defined a knowledge cluster as a group of agents whose mutual distances are smaller than the threshold interaction radius, and whose distance with every node outside the cluster is larger than this radius (meaning that all the agents in the cluster will asymptotically converge to one attractor and no further inclusion of any other agent in the cluster is possible).
We have found that the number of knowledge clusters observed at the end of the network evolution decreases by increasing the threshold interaction radius, because the agents are able to collaborate with partners located farther away in the knowledge space, thus converging all together towards one position.

When the knowledge regime is strongly tacit or strongly explicit, the number of knowledge clusters depends only on the interaction radius itself, and not on the alliance rewiring rate.
The most interesting case occurs for intermediate knowledge regimes, in which the number of knowledge clusters increases with the rewiring rate.
Small rewiring rates lead to the emergence of only one knowledge cluster, which is dispersed in the knowledge space and does not clearly exhibit the presence of a knowledge attractor.
Faster alliance rewiring rates allow the emergence of a larger number of knowledge clusters; in this case, the presence of knowledge attractors, around which the firms eventually cluster, is (even visually) clear.
In such a regime, the agents travel -- on average -- longer distances in the knowledge space. However, if the rewiring rate is too high, the effect of having more knowledge attractors in the system is detrimental for the agents, which do eventually explore a shorter distance.

\paragraph{Our interpretation.}
%to potentially have collaborations with more partners
The underlying assumption of our agent-based model is that the exploration of as many locations as possible is beneficial for the entire collaboration network.
For this reason, we consider the distance explored by the agents in the knowledge space as a \textit{performance indicator} of the network evolution.
We have found that there exists an inverted U-shaped dependence of such an indicator on both the alliance rewiring rate and the interaction radius.

In particular, if we focus on the dependence of the performance on the rewiring rate, as already mentioned, we find that there exists a specific value of the rewiring rate maximizing the performance. Such a rate exhibits a weak dependence on the interaction radius; namely, it slightly decreases when the radius increases (only for intermediate radius values). This is consistent with some empirical studies \cite{rosenkopf2007comparing,gulati2012:_rise_and_fall_of_small_world}, that show a varying alliance formation rate across industrial sectors. 
Similarly, we have found that, given a fixed alliance rewiring rate, there exists a value of the interaction radius maximizing the network performance.

From the point of view of our agent-based model, this happens because in a regime dominated by the exploratory search for new alliance partners, a high rewiring rate allows the agents to cut their links and form new ones before the approaching mechanism with the previous partner is complete, thus traveling with a peculiar ``zig zag'', meandering trajectory.
This has a beneficial effect on the system's performance, because such peculiar trajectories result in a longer traveled distance in the knowledge space.

However, when the rewiring rate is too high, the longer meandering trajectories are no longer able to compensate for the shorter distances globally traveled by the agents, due to the emergence of too many attractors (the agents have to travel less, on average, before converging to one of them). As we have already detected, this results in an inverted U-shaped dependence of the network performance on the rewiring rate.

The same dependence of the network performance occurs as a function of the interaction radius. A higher radius allows the agents to form links even with agents located very far away in the knowledge space, causing a higher distance to be traveled, on average.
A too high radius, on the other hand, causes the emergence of fewer attractors, because the agents can interact with more potential partners, thus converging faster to one or few attractors.
This results in a lower distance globally traveled by the agents to reach the one (or few) attractors, as opposed to the longer distance that they would travel to reach the many different attractors in a scenario with medium interaction radius (and medium/high rewiring rate).

\paragraph{Conclusions.}
In conclusion, the present agent-based model has allowed us to understand how a set of collaborating agents can better explore the knowledge space in which they are located. Our model, at the same time, has shed light on two important aspects of R\&D networks, that can possibly be extended to other collaboration networks: the optimization of the network performance in terms of knowledge exploration, and the emergence of clusters in the knowledge space where the agents interact.
Our results, combined with the empirical observation of different alliance formation rates in different industrial sectors \cite{rosenkopf2007comparing,tomasello2013riseandfall}, could be considered as the first step towards the empirical validation of the performance of R\&D alliance networks.

However, more ingredients can be added to the model in order to capture further effects observed on real R\&D networks or other kinds of collaboration networks.
The first possible extension is the inclusion of more complex strategic link formation rules between the agents, together with the relaxation of the monogamous network approximation, similarly to previous works \cite{konig2011network,koenig2013nestedness,tomasello2014therole}.
In this way, one could investigate network topologies that are closer to the empirical observations.
The alliance formation rules might also be extended with the addition of an inverted U-shaped -- rather than linear -- relationship between the success of a link activation and the knowledge distance of the two involved partners.

Further extensions consist in the addition of a stochastic term to the agents' motion in the knowledge space, to model the firm self-innovation dynamics, or by the adoption of an open-ended knowledge space, that could be more realistic in high-technology industrial sectors.
Another extension is represented by the study of different indicators of the network performance; for instance, one could analyze the share of the knowledge space that has actually been explored by the agents, as opposed to the distance traveled.
A more conceptual research question is whether any measure of traveled distance is a better performance indicator than the number of emerging knowledge attractors, which represents instead a more static, equilibrium measure. One could probably take both kinds of indicators into account, or combine them in an appropriate manner; this would surely require a case-by-case discussion, depending on the system under examination. 

However, provided that appropriate methodologies are known to locate the interacting agents in a metric knowledge space, this model paves the way for further empirical studies not only on R\&D networks, but also on collaboration networks in general.
The scope would be to measure knowledge positions and trajectories of agents in real knowledge spaces, using -- just to name two prominent examples -- patent data for firms, or publication data for scientific authors.
In the case of empirical R\&D networks, alliance formation rates and knowledge regimes characterizing a set of industrial sectors could be quantified and compared, allowing for a check of the consistency of our model with the observed variations in alliance activities across sectors.

\section*{Acknowledgements}
The authors wish to acknowledge the anonymous reviewers, whose comments have greatly improved the quality of this study, in particular the interpretation of our findings. M.~V.~T. and F.~S. acknowledge financial support from the Swiss National Science Foundation (SNF) through grant 100014\_126865, ``R\&D Network Life Cycles'', and from the ETH Zurich Risk Center Seed Project SP-RC 01-15, ``Performance and resilience of collaboration networks''.

\bibliographystyle{ws-acs}
%\bibliography{all_references_MT.bib}

\end{document}